\documentclass[12pt,english,russian]{article}

\usepackage{babel}
\usepackage{varioref}
\usepackage[pt154]{inputenc}
\usepackage[T2A]{fontenc}

\begin{document} 

\begin{center} 
{\Large\bf A note on the mechanism } \\ [1mm]
{\Large\bf of substitution of labour with capital }\\ [1mm]
{\Large\bf in the production processes }

\vspace{5mm}

{\large\it Vladimir Pokrovskii}\footnote{Corresponding author:  Vladimir Pokrovskii, vpok@comtv.ru} \\

\vspace{3mm}


Moscow State University of Economics, Statistics and Informatics, \\ Moscow, RUSSIA 119501

\vspace{5mm}

\end{center}

\centerline{Abstract}

Considering the production processes, it was noted that the use of various equipment leads to an increase in output --- the phenomenon that is usually described as the  substitution of labor with capital. The proposed theory of substitution is based on the assumption that not the quantity of capital (production equipment) does substitute labor, but rather its ability to operate similar to the  workers. This is the true content of the substitution of labor by capital. To formulate a correct mechanism of substitution requires considering three factors of production: the amount of production equipment (capital $K$), human activity (labor $L$), and the substitutive capacity of equipment (substitutive work $P$). The technological properties of production equipment are characterized by the technological coefficients $\overline \lambda$ and $\overline \varepsilon$, indicating the amount of labor and energy required to engaged with  a unit of equipment. The production function can assume various forms, none of which coincide with the popular Cobb-Douglas expression, which seems to be erroneous in its core. 

 \vspace{5mm}
\vspace{5mm}





\vspace{5mm}


\newpage

\section{Introduction}
                                             
Considering the production processes, it was noted that the growth rate of value production could exceed the growth rate of labor efforts, indicating the need for an additional source of added value. To describe this phenomenon, one has been introducing the concept of the productive power of production equipment (called physical capital) into the theory so that the expression for the production of value $Y$ can be written in terms of two factors: labor efforts $L$, measured, for example, in working hours per year, and capital $K$, as the value of production equipment measured in monetary terms, 
\begin{equation}
Y = Y (K, L). 
\end{equation} 
This relationship formalizes the phenomenon of substitution:  the observation that the same quantity of value can be obtained with different amounts of capital (production equipment) and labor efforts. It does not imply any constant proportion for substitution of production factors. 

The researchers were disappointed when they found that the specification of the production function (1) in the assumed  homogeneous form does not quantitatively describe the observed phenomena of economic growth (value production). To achieve correspondence with empirical observation, Solow (1956) suggested that the arguments in the production function are not capital and labor efforts, but associated capital and labor services $A_K(t)K$ and $A_L(t)L$. In one of the simplest and most common cases, the production function can be written in the Cobb-Douglas form, with an additional empirical multiplier $A(t)$, which is supposed to describe exogenous technical progress, 
\begin{equation}
Y = Y_0 A(t)\, \frac{L}{L_0}\left(\frac{L_0}{L} \frac{K}{K_0}\right)^a, \quad 0 < a < 1.
\end{equation} 
With the appropriate choice of multiplier $A(t)$ and index $a$, the production function (2) allows investigators to describe the change in output over time. The time-dependent multiplier $A(t)$ in formula (2) remains unexplained, despite various interpretations and modifications of the production function (Aghion and Howitt,1998, 2009). In this respect, it is unclear what is associated 
with technological progress and clearing this out is known as a challenge of endogenous technical progress. 

To try to find the elusive endogenous technological progress, one could turn to the functional role of equipment in the production process, which is, according to many researchers,  the replacement of human efforts by the work of machines powered by external energy sources. This is consistent with thermodynamic representation of production processes. Machines and devices are installed to perform special operations and facilitate specific work, and the quantity of machines is not as important as the benefits of the equipment installed. The efforts of workers are replaced by the work, called substitutive work, performed by production equipment. Note that this quantity should be considered an independent variable ---  a service provided by production equipment (capital). These premises allow us to understand the substitution mechanism, which includes endogenous technological progress.

\section{The mechanism of substitution}

\subsection{Technological characteristics of capital} 

Production equipment, usually referred to as physical capital $K$, represents the material realization of technology and is used to engage labor $L$ and substitutive work $P$ in production, such that one can introduce, according to previous speculations (Pokrovski, 1999), the main characteristics of capital, that is labor and energy requirements, 
\begin{equation}
\lambda =  \frac{\ d \,  L}{\ d \,  K},  \quad  \varepsilon = \frac{\ d \, P}{\ d \,  K}.		
\end{equation}
These quantities indicate the amounts of labor and productive energy required to power a unit of production equipment. It is convenient to deal with non dimensional technological coefficients 
\begin{equation}
\overline \lambda (t) = \frac{K}{L}\,  \lambda , \qquad 
\overline \varepsilon (t) = \frac{K}{P}\, \varepsilon.
\end{equation}
These quantities characterize the basic technological level of production equipment. If the quantities are less than unity, it means that labor-saving or energy saving technologies are used at that point in time. In the stationary case, the quantities are equal to unity. 

The variable $K$ describes the overall quantity of production equipment. The technological coefficients $\overline \lambda$ and $\overline \varepsilon$ appear to say about its quality. We assume that all three quantities are given as characteristics of capital. Based on these quantities, amounts of necessary production factors can be calculated, 
\begin{equation} 
\frac{1}{L}\,\frac{d L}{d K} = \frac{1}{K}\,\overline \lambda, \qquad \frac{1}{P}\,\frac{d P}{d K} = \frac{1}{K}\,\overline \varepsilon.
\end{equation}

The overall level of production is characterized by the technological index, which is the combination of dimensionless technological coefficients, 
\begin{equation}
\alpha = \frac{1 - \overline{\lambda}}{\overline{\varepsilon} - \overline{\lambda}}.
\end{equation}	
The technological coefficients and time-dependent index ---  the technological index ---  are internal parameters of the production system. 

\subsection{Specification of the production function} 

In a simple approximation, a production system can be represented by  fixed capital (a collection of equipment measured by its value $K$), which acquires the ability to operate through labor and capital services, that is, human efforts $L$ and machine work $P$, with the help of natural energy sources (wind, water, coal, etc.). The work that needs to be done can be performed directly by workers or by devices that use external energy sources to perform the same work. 

Though production equipment is an essential part of a modern enterprise, it remains dead until the efforts of workers and the operation of production equipment are applied to it. The process of production is associated with the activity of people and the work of production equipment. Therefore, the market value of products created is expressed in terms of relationships, 
\begin{equation}
Y = Y [L(K), P(K)], \quad Y = Y(K).		
\end{equation}
The general form of the dependence of output on production factors can be specified on the basis of empirical observations and simple reasoning. For small values of substitutive work, $L >> P$, the substitution effect can be neglected, and the theory reduces to the labor theory of value. The substitution effect is noticeable when the external work significantly overpasses 
the assessment of labor efforts, $P >> L$, while the energy $P$ and labor expenditure $L$ are interchangeable. The amount of production equipment, 
universally measured by its value $K$, should be considered complementary to the work of labor and the production equipment, $L$ and $P$.

 This and some other requirements for the production function (details can be found in the author's monograph: Pokrovskii, 2018), allow us to write the expression for the production of value of in the form 
\begin{equation} 
Y = Y_0 \, \frac{L}{L_0}\left(\frac {L_0}{L} 
\frac{P}{P_0}\right)^\alpha, \quad 0 < \alpha < 1,
\end{equation}
It is assumed that people's activities and the substitutive work of production equipment are equivalent to each other in all aspects. Note that the index $\alpha$ in equation (8) is the same technological index defined by equation (6). This relationship (8) demonstrates that the level of technological development as an internal characteristic of the production system is included in the production function. 

Relations (5) and (6) make it possible to represent the production function (8) as a function of the arguments $L$ and $K$, which are usually used in interpreting substitution effects, 
  \begin{equation}
Y = Y_0 \, \frac{L}{L_0}\left(\frac{ K }{ K_0}\right)^{( 1 -  \overline \lambda)}, \quad 0 < \overline \lambda < 1. 
\end{equation}
This relation can replace expression (2) when estimating the production of value based on the factors $L$ and $K$. Unlike empirical equation (2), relations (8) and (9) are a consequence of a thermodynamically correct substitution mechanism that takes into account the technological characteristics of the equipment.

Note that, by virtue of the validity of equations (5), the functions (8) and (9) can be reduced to a simple relation 
  \begin{equation}
Y = Y_0 \, \frac{ K }{ K_0}. 
\end{equation}

Expressions (8)-(10) represent three forms of the production function; formula (9), for the  given labor requirenent  $\overline\lambda$, determines  the amount of capital needed to replace the unit of labor (elasticity coefficient). Unlike the arguments about the function (2), the theory does not require the introduction of any arbitrary values to be consistent with empirical observations. This situation has been discussed in detais l in the author's previous works (Pokrovskii, 2018, 2021).

An increase in the volume and value of production capital $K$ leads to an increase in output $Y$ , and this observation led to the statement on the productive power of capital. The statement that capital determines the increase in value, but the only  true sources of value are the labor efforts $L$ and the 
work of external energy sources $P$, which are interchangeable and equivalent. In this sense, labor remains the only universal and accurate measure of value, or the only standard by which we compare the values of goods at all times and in all places. The Smith-Marx theory on labor value is complemented by the law of substitution. This law states that in production, the work of third-party forces through production equipment substitutes for human labor. Labor operates  as labor plus equipment. 

\subsection{The marginal productivity of capital}  

The contribution of production factors to the production value is characterized by the marginal productivity of production factors. Due to relation (8), one can easily calculate the marginal productivity of active factors $L$ and $P$, 
\begin{equation}
\beta = \frac{\partial Y}{\partial L}, \quad \gamma  = \frac{\partial Y}{\partial P}. 		
\end{equation}
Human efforts is applied directly, while substitutive work originates from prior exploration of natural laws, development of technological systems, and manufacturing of production machinery, among other things. 

The use of huge amounts of energy in production is impossible without availability of production equipment, which is measured in terms of its value as $K$. This quantity is also known as physical production capital and should be considered as a passive factor of production, unlike factors $L$ and $P$. Physical capital itself is not a source of value but, due to its functional role in production, the "marginal productivity of capital" \, can be formally calculated. Referring to relations (7), we find
\begin{equation}
\frac{d \, Y}{d \,K} = \beta \frac{\ d \,  L}{\ d \,  K} +  \gamma  \frac{\ d \, P}{\ d \,  K}		
\end{equation}

This relationship shows that the productivity of capital (production equipment) is determined by marginal productivities (11) of active production factors and technological coefficients (3). 

The relation (12) demystifies the idea of the productive power of capital. An analysis of the substitution mechanism shows that the efforts of workers to produce goods are replaced by the work of production equipment, not by the equipment itself. The mystical power of capital to bring income arises as a result of the established rules for the distribution of the social product created by  labor and the substitutive work. 

In the public consciousness, the myth of the productivity of capital has ingrained, and this applies not only to productive capital, but to capital in all forms. If we own shares in enterprises, we receive dividends; if our money is in the bank, we earn interest. Stocks and cash are capital in its broadest sense. However, cash and shares are just symbols that do not bring anything without a significant amount of work to produce value within the capitalist system of the national economy.

\section{The concluding remarks }

The uncovering of the mechanism of substitution allows us to formulate a production function, which has three equivalent forms (8) - (10). To explain and describe the phenomenon of enhancement of labor productivity through the introduction of machine production, one needs to consider three production 
factors: capital ($K$), labor ($L$) and substitutive work ($P$). The theory demonstrates that, when describing the production of value, it is the technological property, rather than the amount of capital, that matters. The technological level of production is characterized by the technological coefficients $\overline \lambda$ and $\overline \varepsilon$, which indicate the amount of labor and productive energy required to power production equipment. 

Note that the theory fails to validate the empirical formula (2). It appears that the widely used expression for output (2) contains incorrect assumptions. The production function (2), while considering the availability of production equipment, neglects any quality attributes that might be linked to technological 
processes. 

And in conclusion, we should stress the basic role of the law of value production; it determines the interpretation of production processes and justifies rules for the distribution of social products; in other words, it has a fundamental role in economics and therefore deserves to be considered a basic economic law. In the end, the question is how to correctly interpret observed social phenomena. 

\newpage

\section*{References}

\begin{enumerate}

{\small 

\item Aghion Ph., Howitt P.W. (1998) Endogenous Growth Theory. The MIT Press, Cambridge, Mass. 

\item Aghion Ph., Howitt P.W. (2009) The Economics of Growth. MIT Press. Cambridge, MA. 

\item Pokrovski V.N. (1999) Physical Principles in the Theory of Economic Growth. Ashgate Publishing, Aldershot. 

\item Pokrovskii V.N. (2018) Econodynamics: The Theory of Social Production, The 3nd Ed.. Springer, Dordrecht-Heidelberg-London-New York, 2018 \\ https://www.springer.com/gp/book/9783319720739 . 

\item Pokrovskii V.N. (2021) Social resources in the theory of economic growth. The Complex Systems, No 3 (40), 33 - 44. Available at Researchgate. 

\item Solow R. (1957) Technical change and the aggregate production function. Review of Economic Studies, v. 39, Aug., p. 312-330. 
}

 \end{enumerate}

\end{document}